\documentclass[twocolumn,showpacs,prb]{revtex4}

\usepackage{graphicx}
\usepackage{dcolumn}
\usepackage{bm}

\begin{document}

\title{Electronic band structure of novel 18-K\\
superconductor Y$_{2}$C$_{3}$ as compared with YC and YC$_{2}$.}

\author{I.R. Shein$^*$ and A.L. Ivanovskii}

\affiliation {Institute of Solid State Chemistry, Ural Branch of
the Russian Academy of Sciences, 620219, Ekaterinburg, Russia}

\begin{abstract}
The electronic band structure of yttrium sesquicarbide
Y$_{2}$C$_{3}$ (Pu$_{2}$C$_{3}$ structural type) reported by
Akimitsu et al. (2003) as a novel 18-K superconductor is
investigated using the first-principle full-potential LMTO method
and compared with those of yttrium mono- and dicarbide: cubic YC
and 4-K superconductor YC$_{2}$ (CaC$_{2}$ structural type). Our
results show that the enhanced T$_{c}$ of Y$_{2}$C$_{3}$ as
compared with YC$_{2}$ may be interpreted by the electronic
factors: the near-Fermi DOS for Y$_{2}$C$_{3}$ was found to be
about 70 $\%$ higher that for YC$_{2}$, and the contribution from
the C2p states increases. The Fermi level of the "ideal" bcc
Y$_{2}$C$_{3}$ is located near the local DOS minimum, and the
superconducting properties of Y$_{2}$C$_{3}$-based materials would
be very sensitive to synthesis conditions and the presence of
impurities and lattice vacancies. We suppose that by changing
these factors it is
possible to vary T$_{c}$ for the yttrium sesquicarbide.\\

$^*$ E-mail: shein@ihim.uran.ru
\end{abstract}



\maketitle

I. Introduction.\\

Metal carbides represent a technologically important group of
materials with interesting properties including
extreme hardness, high melting temperatures, and superconductivity \cite{Toth}.\\
Among superconducting binary compounds, metal carbides are known
as conventional low-temperature BCS superconductors (SC) and have
long been examined \cite{Vonsovsky}. Comparison of different
classes of binary carbides: semi- (M$_{2}$C), mono- (MC), sesqui-
(M$_{2}$C$_{3}$), and dicarbides (MC$_{2}$) shows (reviews
\cite{Vonsovsky,Poople}) that the majority of the known
superconductors are found among cubic compounds with a rock-salt
structure (B1), in which C and M atoms are in octahedral
surrounding. Among these phases, the maximum values of transition
temperatures T$_{c}$ belong to Ta (10.2), W (10 K), Nb (11.1), and
Mo (14.3 K) monocarbides \cite{Poople}. Using the first-principle
band structure calculations results, it was proposed that B1-MoC
(isoelectronic and isostructural to the 16.6-K superconductor NbN
\cite{Poople}) was the prime candidate for the
"highest-temperature" SC among all familiar metal carbides with a
predicted transition temperature of about 20 K \cite{Hart}.\\
The superconducting state is much less pronounced for the
dicarbides, in which C atoms are in the form of isolated C$_{2}$
pairs. Among metal dicarbides, the low-temperature SC was observed
only for three compounds MC$_{2}$ (M = Y, La, Lu) with the highest
T$_{c}$ value of 3.9 K for YC$_{2}$ \cite{Poople}. In addition,
some superconducting metal sesquicarbides have been also
described, for example several compounds with T$_{c}$ of about 11
K (La$_{2}$C$_{2.7}$ and Y$_{2}$C$_{3}$) and 15 K
(Lu$_{2}$C$_{3}$) \cite{Poople}.\\
The interest in the properties of the yttrium carbide systems
increased in the past years due to the discovery of
superconductivity with transition temperatures up to 11.6 K in the
layered yttrium carbide halides Y$_{2}$C$_{2}$I$_{2}$ (T$_{c}$ =
9.97 K) and Y$_{2}$C$_{2}$Br$_{2}$ (T$_{c}$ = 5.04 K) \cite{Henn}
and also because of the observation of superconductivity with
T$_{c}$ $\sim$ 15-23 K \cite{Cava,Cava1} in more complex systems:
Y-Ni and Y-Pd-based borocarbides, see also reviews
\cite{Ivanovskii,Behera,Yang}. In this context, binary and
quasi-binary Y carbides (especially Th-doped yttrium dicarbide
(Y$_{1-x}$Th$_{x}$)C$_{2}$ \cite{Krupka}) also attracted
particular attention since their T$_{c}$ values are
close to those of A15 compounds \cite{Vonsovsky,Poople}.\\
Quite recently, Akimitsu et al \cite{Amano} have carried out a
high-pressure synthesis of a novel superconducting phase in the
Y-C system with T$_{c}$ of about 18 K and assumed that this 18-K
SC originated in the yttrium sesquicarbide Y$_{2}$C$_{3}$ with the
body-centered cubic (bcc) Pu$_{2}$C$_{3}$-type crystal structure.\\
The value of T$_{c}$ for Y$_{2}$C$_{3}$ obtained in magnetic
measurements \cite{Nakane} was reported earlier to be 11 K. It was
pointed out \cite{Krupka} that the superconductivity in the Y-C
yttrium system is very sensitive to the synthesis conditions, for
instance pressure and heat treatment. In particular, the effect of
carbon nonstoichiometry was observed. In Ref.\cite{Nakane}, the
bcc Y$_{2}$C$_{3}$ has been synthesized by arc-melting and
high-pressure techniques. Magnetic susceptibility showed that the
material exhibited superconductivity with variable T$_{c}$ ranging
from 6 K to 11.5 K. In Ref.\cite{Amano}, the superconducting
Y$_{2}$C$_{3}$ phase was prepared under high pressure using mixed
powders (Y+C) in a BN cell. It is noteworthy that the lattice
parameter (a) changes for different sintering processes and exists
in the range between 0.8181 and 0.8226 nm \cite{Amano} as distinct
from earlier findings (0.8214 - 0.8251 nm) \cite{Krupka}. One of
the reasons may be the difference in the pressure range (4-5.5
\cite{Amano} compared with 1.5-2.5 GPa \cite{Krupka}).
Furthermore, the authors \cite{Nakane} successfully reproduced the
result \cite{Amano} and reported the synthesis of the 18-K phase
with some differences in the experimental technique. In addition,
a sample with the nominal composition Y$_{2}$C$_{2.9}$B$_{0.1}$
was prepared to check a possible role of boron impurities from the
BN cell used in the experiments \cite{Amano}. According to data
\cite{Nakane}, the 18-K phase coexists with another phase (the
11-K phase reported in \cite{Krupka}) having a trace of a
low-T$_{c}$ phase (4 K). This "impurity" phase is YC$_{2}$.
Moreover, the T$_{c}$ for the boron-doped sample (16.4 K) is lower
than that of the "pure" carbide. This means that boron impurities
in the C positions play a negative role in the Y-C
superconductivity. It was also reported that the magnetization
(M-H) curves of the 18-K Y$_{2}$C$_{3}$ phase showed a type-II
superconducting behavior \cite{Amano}. While an explanation for
the observed \cite{Amano,Nakane} behavior in the 18-K material
based on a microscopic theory is still lacking, it is necessary to
perform comparative studies of the electronic properties of
yttrium carbide phases. In this paper, we report the theoretical
results related to the electronic band structure of bcc
Y$_{2}$C$_{3}$ in comparison with other yttrium carbide phases,
namely the B1 monocarbide YC and 4-K SC YC$_{2}$. The electronic
bands, density of states (DOS) and site-projected l-decomposed DOS
near the Fermi energy (E$_{F}$) of these carbides are  obtained
and analyzed as a function of structures and nominal stoichiometry:
 YC (C/Y=1) $>$ Y$_{2}$C$_{3}$ (C/Y=1.5) $>$ YC$_{2}$ (C/Y=2).\\

II. Models and Method of Calculation.\\

The yttrium carbides have cubic (YC, Y$_{2}$C$_{3}$) and
tetragonal (YC$_{2}$ - space group I4/mmm) crystal structures. The
yttrium sesquicarbide crystallizes in a bcc structure of the
Pu$_{2}$O$_{3}$ type (space group I-43d) with the lattice
parameter a = 0.8233 nm (ICSD-CC No.86290) and eight formula units
per a unit cell\cite{Novokshonov}. The Wyckoff positions for atoms
are Y:16c (0.05;0.05;0.05) and C:24d (0.2821;0;1/4). The band
structure of the above-mentioned carbides was calculated using the
scalar relativistic full-potential linear muffin-tin method
(FLMTO) \cite{Savrasov} with the generalized gradient
approximation (GGA) of Perdew et al. \cite{Perdew}.The
Y$_{2}$C$_{3}$ structural parameters are taken from Ref.
\cite{Novokshonov}. The equilibrium values of lattice constants
obtained from minimum total energy and are a = 0.5076 nm
 for YC and a = 0.3699 nm, c = 0.6140 nm, x = 0.3748 for YC$_{2}$.\\

III. Results.\\

In Fig. 1 we show our FLMTO band structure of B1-YC. The YC band
structure is very similar to that found for other rock-salt (B1)
transition metal carbides, see for example
\cite{Hart,Neckel,Marksteiner,Gubanov}, consist from low-energy
C2s band, does not contribute to the bonding. This band exhibits
the maximum dispersion (2.35 eV) between $\Gamma$ and L points.
The next bands separated from the C2s band by a gap at about 3.34
eV originate mainly from mixed C2p-Y4d states. The Y4d bands are
decomposed into e$_{g}$ and t$_{2g}$ components with energies 1.82
and 3.49 eV in $\Gamma$ point above the Fermi level (E$_{F}$). In
B1 carbides, the e$_{g}$-like bands mixed with the C2p states are
usually viewed as the bonding part of the d-p hybridization zone,
while t$_{2g}$ - as the antibonding part overlapping the carbon 2p
states \cite{Hart,Neckel,Marksteiner,Gubanov}. The phase
instability of stoichiometric YC may be explained based on band
filling \cite{Gubanov}. In YC, the Fermi level intersects the
e$_{g}$-like bands. As a result, the bonding states remain
partially unoccupied as distinct for example from the stable
B1-ZrC phase, where all bonding states are occupied, whereas
antibonding states, on the contrary, are unfilled. The density of
states (per formula unit) at the Fermi level N(E$_{F}$) is quite
large (Table 1) and comparable with N(E$_{F}$) of B1-NbC. At the
same time their compositions are different: for YC the
contributions from d and C2p states are comparable (Table 1),
while the main role in NbC is played by metallic d states. Note
also that inter-atomic bonds and charge distribution in B1-YC are
quite isotropic, see Fig. 2. The electronic properties of yttrium
dicarbide were studied earlier using the tight-binding linear
muffin-tin orbital atomic-sphere approximation (TB-LMTO-ASA)
calculations \cite{Gulden,Zhukov}. Our FLMTO YC$_{2}$ electronic
bands and DOSs are shown in Figs. 3, 4 and agree well with those
reported in \cite{Gulden,Zhukov}. The two lowest bands (Fig. 3) at
about 14.4 and 7.1 eV below E$_{F}$ are bonding and nonbonding C2s
states respectively. The low-lying bonding C2s band with the
maximal dispersion of ~ 1.1 eV is separated from the next band
group by a gap of about 6.4 eV. The combination of the C2p$_{x,y}$
orbitals forms the quasi-flat bands from $\Gamma$ to Z with the
DOS maximum around -4.2 eV, which are hybridized with the
Yd$_{xy}$ orbitals, Fig. 4. The C2p$_{z}$ states split up into two
bands (peaks A, B in the partial DOS picture around -7.3 and -2.2
eV) hybridized with the Yd$_{3z^2-1}$ states. All these bands
contribute to the Y-C metal bonding and are completely occupied.
The partially filled near-Fermi bands (Y4$_{yz,xz}$ in the region
-2.0 eV $\div$ E$_{F}$ and Y4$_{x^2-y^2}$ in the region -0.95 eV
$\div$ E$_{F}$) are strongly dispersive and are hybridized with
the antibonding C2p$_{x,y}$ orbitals in the direction of axis c.
The Fermi level intersects three bands near $\Gamma$, P, and Z
points. These bands form the Fermi surface, which contains, in
particular, hole sheets (around points $\Gamma$ and Z), Fig. 5.
The composition of N(E$_{F}$) includes the contributions from the
Yd$_{xz,yz,x^2-y^2}$, and C2p$_{x,y}$ orbitals, see Table 1. The
value of the other Yd, C2p components, as well as of the Yp,s and
C2s valence states are negligibly small. Thus the electronic
structure and the chemical bonding type in yttrium mono- and
dicarbide differ essentially: isotropic Y-C bonds in YC versus
anisotropic Y-C and C-C bonds in YC$_{2}$, see charge-density
contour maps in Fig.2.\\
The calculated bands and DOS for Y$_{2}$C$_{3}$ are shown in Figs.
6, 7. The valence zone of Y$_{2}$C$_{3}$ contains four separated
band groups (I-IV, Fig. 7). Bands (I) between -14.3 and -13.3 eV
are bonding carbon 2s states, whereas antibonding C2s states (with
an admixture of C2p$_{x}$ and Yd,p,s orbitals) form separate bands
(II) in the region -7.1 $\div$ -6.4 eV. The higher completely
occupied band group (III) between -5.7 and - 2.4 eV comes from
Yp,d and C2p orbitals. The Fermi level lies in the middle of bands
(IV) composed mostly of Yd-C2p states. The bands intersecting the
Fermi level form hole (around H and N points) and electronic sheets
of the Fermi surface.\\
As in YC$_{2}$, the partial DOSs of Y$_{2}$C$_{3}$ demonstrated
very anisotropic orbital states distributions (see also Table 1)
depending on the non-equivalency of Y-C and C-C bonds (Fig. 8) and
the structural peculiarities of the yttrium susquicarbide, which
contains carbon pairs in different orientations
\cite{Novokshonov}. Of special interest is a near-Fermi DOS
Y$_{2}$C$_{3}$, since in the conventional BCS-type superconductors
the DOS within the interval E$_{F}$ $\pm$ h$\Theta_{D}$
($\Theta_{D}$ - Debye frequency) at the Fermi level is crucial for
superconductivity, we show in Fig. 9 the density of states for
Y$_{2}$C$_{3}$ within a small energy interval around E$_{F}$. As
is seen from Fig. 9, E$_{F}$ is in the region of DOS minimum
between two sharp peaks A and B and falls to a local small DOS
peak. Thus, both electronic and hole doping that change the
position of E$_{F}$ may decrease or increase N(E$_{F}$) and
accordingly T$_{c}$.\\
Simple estimations based on the rigid-band model show that E$_{F}$
will coincide with maxima A and B, when about 0.7 holes or 0.5
additional electrons are introduced per formula unit of
Y$_{2}$C$_{3}$. As a result, N(E$_{F}$) of these doped systems
increases almost twice (by 89 and 98 $\%$ respectively in
comparison with N(E$_{F}$) of the stociometric yttrium
sesquicarbide). This effect may be achieved by several means. The
most probable approach is to dope the yttrium sublattice with
donor or acceptor dopants. The role of hole dopants may be played
also by carbon vacancies. Another important factor N(E$_{F}$) of
the metastable Y$_{2}$C$_{3}$ phase may be pressure
\cite{Gulden,Amano,Krupka}. The immediate prospects of our studies
will involve simulation of pressure effects on fine peculiarities
of the electronic band structure of Y$_{2}$C$_{3}$. In summary,
our FLMTO calculations of the electronic band structure of the bcc
yttrium sesquicarbide indicate that the enhanced transition
temperature as compared with YC$_{2}$ may be due to electronic
factors: the DOS at the Fermi level for Y$_{2}$C$_{3}$ was found
to be about 40 $\%$ higher than that for 4-K SC YC$_{2}$, Table 1.
In addition, the contribution from the C2p states increases
appreciably. Since the Fermi level of the "ideal" bcc yttrium
sesquicarbide is located near the local DOS minimum, the
superconducting properties of Y$_{2}$C$_{3}$-based materials will
be very sensitive to synthesis conditions (for example, pressure
and annealing effects) and the presence of lattice defects. By
changing these factors it is possible to vary the
T$_{c}$ of the Y-C phase in a wide range.\\

Acknowledgment.\\

This work was supported by the RFBR, grant 02-03-32971, and the
Russian Foundation for Scientific Schools, grant SS 829.2003.3.


\begin{table}
\caption{. Total and site-projected $\ell$-decomposed DOSs at the
Fermi level (N(E$_F$), states/eV).}
\begin{center}
\begin{tabular}{|c|c|c|c|}
\hline
Density of state & & &\\
Partial states on atom &YC&YC$_{1.5}$&YC$_{2}$\\
Total state on molecule & & & \\
\hline
{\bf C2s}&0,0088&0,0076&0,0104\\
\hline
C2p$_{x}$&0,3849&0,1856&0,0501\\
C2p$_{y}$&0,3849&0,0293&0,0501\\
C2p$_{z}$&0,3849&0,0937&0,0089\\
\hline
 {\bf C2p}&1,1546&0,3086&0,1091\\
\hline
{\bf Ys}&0,0084&0,0260&0,0039\\
\hline
Yp$_{x}$&0,0221&0,0153&0,0004\\
Yp$_{y}$&0,0221&0,0153&0,0004\\
Yp$_{z}$&0,0221&0,0153&0,0004\\
\hline
 {\bf Yp}&0,0664&0,0459&0,0011\\
\hline
Yd$_{yz}$&0,0918&0,0843&0,2048\\
Yd$_{xz}$&0,0918&0,0843&0,2048\\
Yd$_{xy}$&0,0918&0,0843&0,0060\\
Yd$_{x^2-y^2}$&0,1064&0,2059&0,1036\\
Yd$_{3z^2-1}$&0,1064&0,2059&0,0025\\
\hline
{\bf Yd}&0,4882&0,6647&0.5217\\
\hline
{\bf Total}&1.7265&1.2107&0.7658\\
\hline
\end{tabular}
\end{center}
\end{table}

\begin{figure*}[!htb]
\vskip  0cm
\begin{tabular}{c}

\includegraphics[width=9.0 cm,clip]{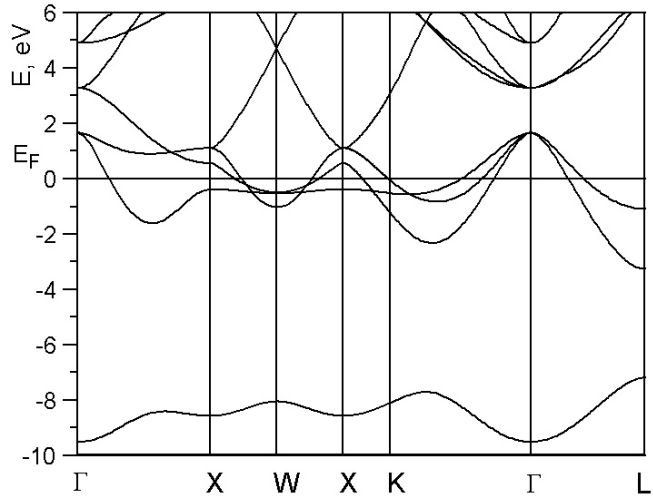} \\

\end{tabular}
\vspace{-0.02cm} \caption[a] { \small The band structure of
hypothetical B1-YC.  } \label{fig:FS}
\end{figure*}

\begin{figure*}[!htb]
\vskip  0cm
\begin{tabular}{cc}
1 & 2   \\

\includegraphics[width=6.2 cm,clip]{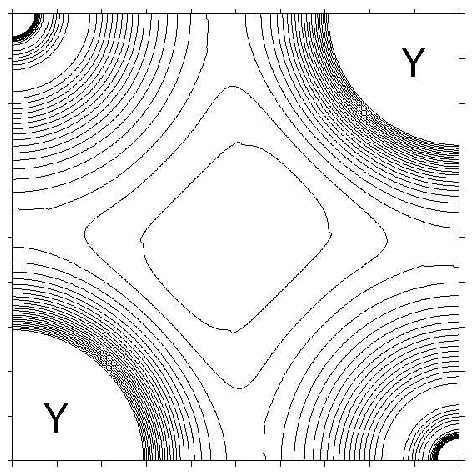}
&
\includegraphics[width=5.2 cm,clip]{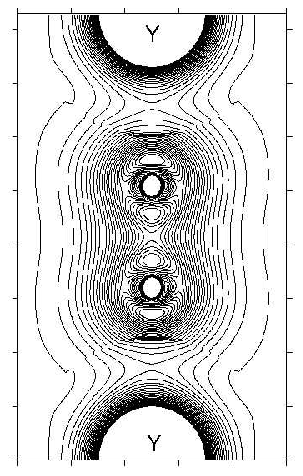}\\

\end{tabular}

\vspace{-0.02cm} \caption[a] { \small Constant charge-density
contour maps in sections of yttrium mono- (1) - and dicarbide (2)
cells illustrating the formation of isotropic Y-C and anisotropic
C-C bonds respectively.} \label{fig:FS}
\end{figure*}

\begin{figure*}[!htb]
\vskip  0cm
\begin{tabular}{c}

\includegraphics[width=9.0 cm,clip]{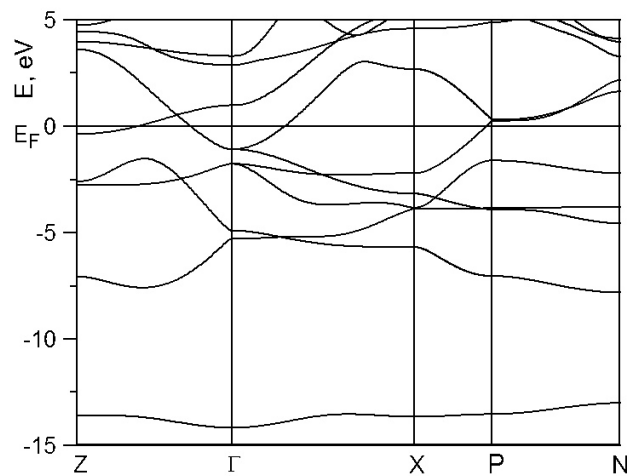}\\

\end{tabular}

\vspace{-0.02cm} \caption[a] { \small The band structure of
YC$_{2}$.} \label{fig:FS}
\end{figure*}

\begin{figure*}[!htb]
\vskip  0cm
\begin{tabular}{c}

\includegraphics[width=12.0 cm,clip]{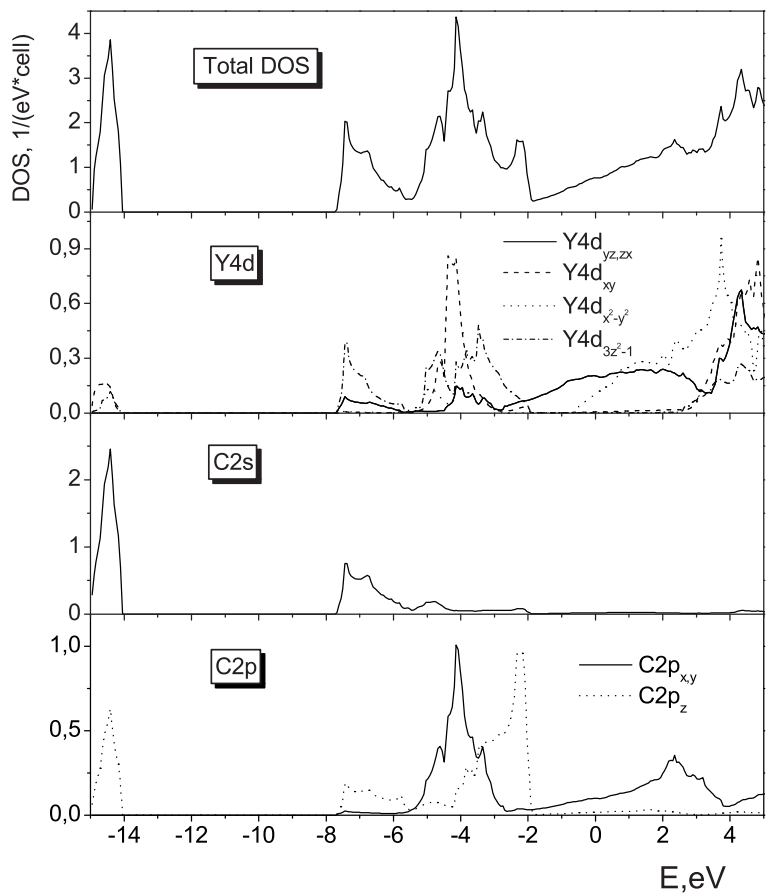}\\

\end{tabular}

\vspace{-0.02cm} \caption[a] { \small The total and partial
densities of states for YC$_{2}$.} \label{fig:FS}
\end{figure*}

\begin{figure*}[!htb]
\vskip  0cm
\begin{tabular}{c}
\includegraphics[width=8.0 cm,clip]{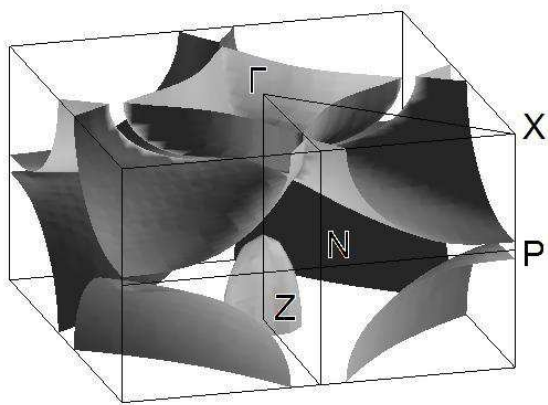}\\
\end{tabular}
\vspace{-0.02cm} \caption[a] { \small The Fermi surface for
YC$_{2}$.} \label{fig:FS}
\end{figure*}

\begin{figure*}[!htb]
\vskip  0cm
\begin{tabular}{c}
\includegraphics[width=9.0 cm,clip]{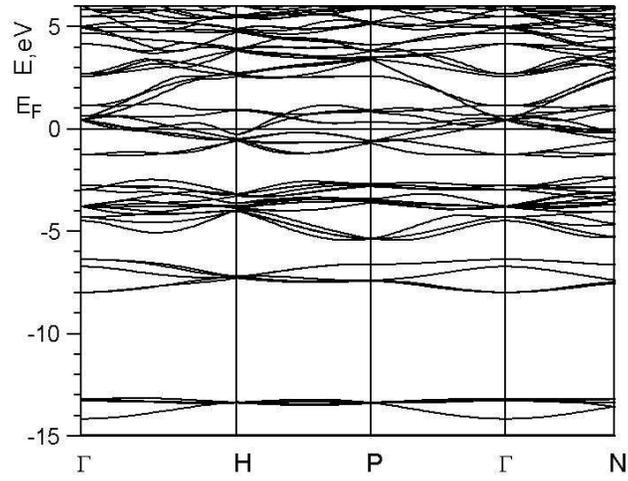}\\
\end{tabular}
\vspace{-0.02cm} \caption[a] { \small The band structure of
Y$_{2}$C$_{3}$ calculated with the lattice constants as given in
Ref.\cite{Novokshonov} using the full-potential LMTO method.}
\label{fig:FS}
\end{figure*}

\begin{figure*}[!htb]
\vskip  0cm
\begin{tabular}{c}
\includegraphics[width=12.0 cm,clip]{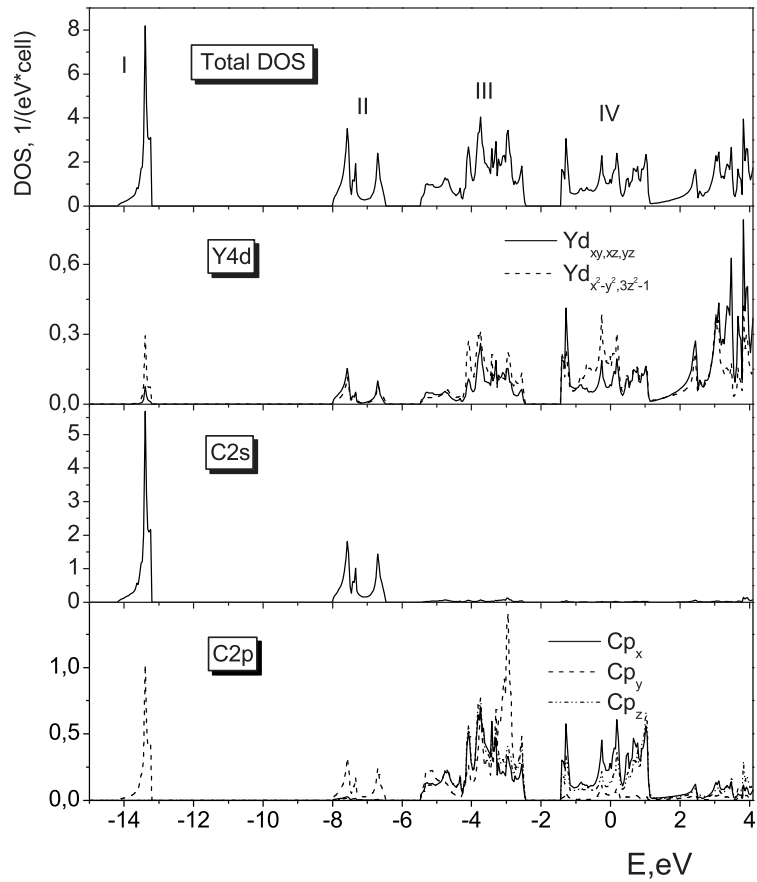}\\
\end{tabular}
\vspace{-0.02cm} \caption[a] { \small The total and partial DOSs
for Y$_{2}$C$_{3}$ calculated with the lattice constants as given
in Ref.\cite{Novokshonov} using the full-potential LMTO
method.}\label{fig:FS}
\end{figure*}

\begin{figure*}[!htb]
\vskip  0cm
\begin{tabular}{c}
\includegraphics[width=5.2 cm,clip]{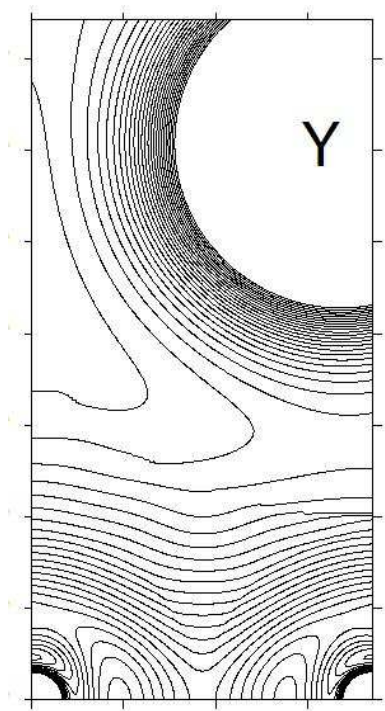}\\
\end{tabular}
\vspace{-0.02cm} \caption[a] { \small The charge density map in
the section of Y$_{2}$C$_{3}$ cell illustrating the C-C covalent
bond. }\label{fig:FS}
\end{figure*}

\begin{figure*}[!htb]
\vskip  0cm
\begin{tabular}{c}
\includegraphics[width=15.0 cm,clip]{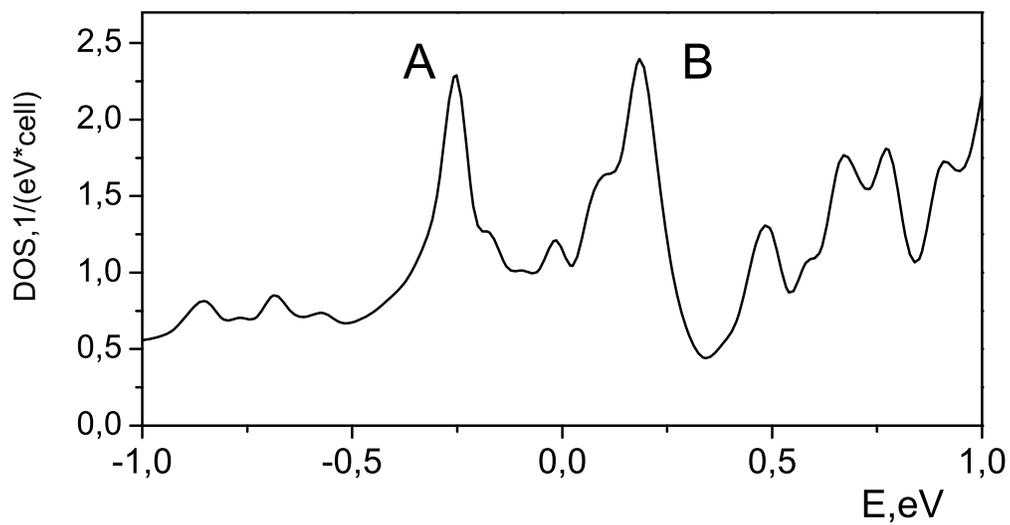}\\
\end{tabular}
\vspace{-0.02cm} \caption[a] { \small The total density of states
of Y$_{2}$C$_{3}$ around the Fermi energy calculated with the
lattice constants as given in Ref.
\cite{Novokshonov}.}\label{fig:FS}
\end{figure*}

\end{document}